\documentclass[linenumbers]{aastex63}

\graphicspath{{./}{figures/}}

\begin{document}
\nolinenumbers

\title{Astrometric Apparent Motion of High-redshift Radio Sources}

\correspondingauthor{Oleg Titov}
\email{Oleg.Titov@ga.gov.au}

\author[0000-0003-1751-676X]{Oleg Titov}
\altaffiliation{}
\affiliation{Geoscience Australia, PO Box 378, Canberra 2601, Australia}

\author[0000-0003-3079-1889]{S\'{a}ndor Frey}
\altaffiliation{}
\affiliation{Konkoly Observatory, ELKH Research Centre for Astronomy and Earth Sciences, Konkoly Thege Mikl\'{o}s \'{u}t 15-17, 1121 Budapest, Hungary}
\affiliation{CSFK, MTA Centre of Excellence, Konkoly Thege Mikl\'{o}s \'{u}t 15-17, 1121 Budapest, Hungary}
\affiliation{Institute of Physics, ELTE E\"otv\"os Lor\'and University, P\'azm\'any P\'eter s\'et\'any 1/A,
H-1117 Budapest, Hungary}

\author[0000-0002-8466-7026]{Alexey Melnikov}
\altaffiliation{}
\affiliation{Institute of Applied Astronomy, Russian Academy of Sciences, Kutuzova Embankment 10, St. Petersburg, 191187, Russia}

\author[0000-0001-7308-6659]{Fengchun Shu}
\altaffiliation{}
\affiliation{Shanghai Astronomical Observatory of the Chinese Academy of Science, 80, Nandan Road, Shanghai, China}

\author[0000-0002-1368-7521]{Bo Xia}
\altaffiliation{}
\affiliation{Shanghai Astronomical Observatory of the Chinese Academy of Science, 80, Nandan Road, Shanghai, China}

\author{Javier Gonz\'{a}lez}
\altaffiliation{}
\affiliation{Observatorio de Yebes (IGN), Apartado 148, 19180 Yebes, Spain}

\author{Bel\'{e}n Tercero}
\altaffiliation{}
\affiliation{Observatorio de Yebes (IGN), Apartado 148, 19180 Yebes, Spain}

\author[0000-0002-0694-2459]{Leonid I. Gurvits}
\altaffiliation{}
\affiliation{Joint Institute of VLBI ERIC, Oude Hoogeveensedijk 4, 7991~PD Dwingeloo, The Netherlands}
\affiliation{Faculty of Aerospace Engineering, Delft University of Technology, Kluyverweg 1, 2629 HS Delft, The Netherlands}

\author[0000-0001-9885-4220]{Aletha de Witt}
\altaffiliation{}
\affiliation{South African Radio Astronomy Observatory, 2 Fir Street, Black River Park, Cape Town, 7925, South Africa
}

\author[0000-0002-0233-6937]{Jamie McCallum}
\altaffiliation{}
\affiliation{University of Tasmania, Hobart, Australia}

\author[0000-0002-0321-8588]{Mikhail Kharinov}
\altaffiliation{}
\affiliation{Institute of Applied Astronomy, Russian Academy of Sciences, Kutuzova Embankment 10, St. Petersburg, 191187, Russia}

\author[0000-0002-1152-4614]{Vladimir Zimovsky}
\altaffiliation{}
\affiliation{Institute of Applied Astronomy, Russian Academy of Sciences, Kutuzova Embankment 10, St. Petersburg, 191187, Russia}

\author[0000-0002-8813-4884]{M\'at\'e Krezinger}
\affiliation{Department of Astronomy, Institute of Geography and Earth Sciences, ELTE E\"otv\"os Lor\'and University, P\'azm\'any P\'eter s\'et\'any 1/A,
H-1117 Budapest, Hungary}
\affiliation{Konkoly Observatory, ELKH Research Centre for Astronomy and Earth Sciences, Konkoly Thege Mikl\'{o}s \'{u}t 15-17, 1121 Budapest, Hungary}



\begin{abstract}
\nolinenumbers

Radio-loud quasars at high redshift ($z \ge 4$) are rare objects in the Universe and rarely observed with Very Long Baseline Interferometry (VLBI). But some of them have flux density sufficiently high for monitoring of their apparent position. The instability of the astrometric positions could be linked to the astrophysical process in the jetted active galactic nuclei in the early Universe. Regular observations of the high-redshift quasars are used for estimating their apparent proper motion over several years. We have undertaken regular VLBI observations of several high-redshift quasars at 2.3~GHz ($S$ band) and 8.4~GHz ($X$ band) with a network of five radio telescopes: 40-m Yebes (Spain), 25-m Sheshan (China), and three 32-m telescopes of the Quasar VLBI Network (Russia) -- Svetloe, Zelenchukskaya, and Badary. Additional facilities joined this network occasionally. The sources have also been observed in three sessions with the European VLBI Network (EVN) in $2018-2019$ and one Long Baseline Array (LBA) experiment in 2018. In addition, several experiments conducted with the Very Long Baseline Array (VLBA) in $2017-2018$ were used to improve the time sampling and the statistics. Based on these 37 astrometric VLBI experiments between 2017 and 2021, we estimated the apparent proper motions of four quasars: 0901$+$697, 1428$+$422, 1508$+$572, and  2101$+$600. 

\end{abstract}



\section{Introduction}

The third version of the International Celestial Reference Frame (ICRF3) is defined by the positions of 4536 extragalactic radio sources \citep{2020A&A...644A.159C} measured by the geodetic Very Long Baseline Interferometry (VLBI) technique in dual-frequency mode ($S/X$), in the framework of the observing campaigns coordinated by the International VLBI Service for Geodesy and Astrometry \citep[IVS,][]{2017JGeod..91..711N}. These extragalactic radio sources are active galactic nuclei (AGN), presumably with a supermassive black hole in the nucleus, mostly appearing in VLBI images as a set of very compact radio-bright knots (a core and jet components). Due to the physical processes in the nucleus area of the sources, the structure of the AGN on milliarcsec (mas) scales probed by VLBI at GHz frequencies is variable on the timescales down to months, leading to either physical motion of the compact knots, or variations of their brightness, or both. As a result, highly accurate positions of the reference radio sources are unstable on milliarcsecond scales.
If a bright jet component is moving persistently along a particular direction for several years, then the effect might be detected as apparent motion of the radio source with a typical magnitude of $\sim 10-100$\,$\mu$as\,yr$^{-1}$ \citep[e.g.][]{1997AJ....114.2284F,2003A&A...403..105F,2007AstL...33..481T,2011AJ....141..178M}. 

High-redshift AGN are tracers of the properties of early cosmological epochs. Thus, the astrometric observing project described in this paper has astrophysical relevance, too. Only a small fraction of high-redshift AGN are strong enough at radio frequencies to be available for VLBI observations using medium-size radio telescopes. Therefore, a multi-epoch astrometry of the high-redshift AGN is a subject of special interest. 

Historically, in astrometry the term ``proper motion'' refers to the physical motion of stars in our Galaxy. This effect is observed as a coordinate shift of the comparatively nearby stars with respect to more distant stars or extragalactic objects. 
As the quasars are at cosmological distances from the Earth-based observer, the corresponding proper motion induced by the same physical reasons is negligible \citep{1986Natur.319..733B}. For instance, even for the closest galaxy with compact radio core (M\,81) at a distance of 3\,Mpc, the proper motion does not exceed $\sim 10$\,$\mu$as\,yr$^{-1}$ \citep{1986SvA....30..501K} and has not been detected so far. The motion of one or more compact components of the AGN jet, or variations of the brightness that lead to large apparent change of the radio source position are not linked to a bulk physical motion of the whole AGN with respect to the observer. Therefore, this astrometric effect is used to be called ``apparent proper motion'' to separate it from the stellar proper motion that have a real physical nature.

Apart from the apparent proper motions induced by the relativistic jets that are randomly distributed in direction for a sample of objects, small systematic effects may also appear. In particular, the dipole systematic of several $\mu$as\,yr$^{-1}$ discloses an impact of the Galactocentric acceleration of the Earth \citep[e.g.][]{ 1997ApJ...485...87G, 1998RvMP...70.1393S, 2011A&A...529A..91T, 2019A&A...630A..93M, 2020A&A...644A.159C}. On the cosmological scale of events, theoretical studies suggest the existence of primordial gravitational waves. The primordial gravitational waves, if they are strong enough to be detected, would cause a quadrupole systematic in the apparent proper motion of the reference radio sources \citep[e.g.][]{1966ApJ...143..379K, 1997ApJ...485...87G, 2011A&A...529A..91T}. Moreover, in some cosmological scenarios, the magnitude of the proper motions could grow with redshift. 
If the cosmological signal is too weak, it might not be detected with the low-redshift radio sources. However, as this signal grows with distance, it might be detected with radio sources at higher redshift. As the number of high-redshift radio sources regularly observed in the geodetic VLBI program is limited, it is not possible to extract a specific systematic signal from a large set of proper motions. The only possible way is to check whether the observed high-redshift sources demonstrate the proper motion that cannot be related to the usual relativistic jet activity.

Therefore, in our observational project, we are focusing on a set of radio sources with redshift $z \ge 4$, trying to detect unusually large proper motions on a timescale of several years. Detection of any astrometric anomaly would trigger a follow-up cosmological research.

\section{The sample and observations}

We started composing the sample from a list of previously observed quasars at redshift $z \ge 4$ with total flux density at 8.4~GHz  $S_{8.4} \ge 100$~mJy. This simple selection criterion gave us four radio sources: 0901$+$697, 1428$+$422, 1508$+$572, and 2101$+$600. The sources' celestial positions favoured regular observations of these quasars by a set of radio telescopes in the northern hemisphere. To achieve a good signal-to-noise ratio (\textit{SNR}) for a fringe detection, we have practiced prolonged tracking of the high-redshift radio sources during the RUA, EVN and LBA experiments, up to 18--20 minutes for selected radio sources.
Other 20--25 radio sources used to tie their positions to the ICRF3 were observed in the standard geodetic mode \citep[e.g.][]{1998RvMP...70.1393S} to achieve \textit{SNR}~$\ge 7$. All the sources were selected from the list of 303~ICRF3 ``defining'' radio sources to provide the best reference frame around the sky with positions known to be better than 30\,$\mu$as.  
The Very Long Baseline Array (VLBA) experiments had used short snapshot scans for all radio sources.

\begin{table}[h]
\caption{List of the high-redshift radio sources discussed in this paper. Positions are from the aus2021b.crf solution (see footnote [\ref{fnote1}]) produced by the Geoscience Australia IVS Analysis Centre. 
}
\begin{tabular}{lcccccll} 
\hline
Source	   &  $z$	& Ref.    	& Right ascension (2000.0) & Declination (2000.0)  & No. of experiments  \\
\hline
0901$+$697  & 5.47  & 1 & 09 06 30.74876   & $+$69 30 30.8287  & 30  \\ 
1428$+$422  & 4.715 & 2 & 14 30 23.74163   & $+$42 04 36.4911  & 25  \\ 
1508$+$572  & 4.301 & 3 & 15 10 02.92237   & $+$57 02 43.3759  & 104 \\ 
2101$+$600  & 4.575 & 4 & 21 02 40.21905   & $+$60 15 09.8366  & 21  \\ 
\hline 
\label{table_sources}
\end{tabular}

\tablecomments{Redshift references: 1 -- \citet{2004ApJ...610L...9R},  2 -- \citet{1998MNRAS.294L...7H}, 3 -- \citet{1995MNRAS.273L..63H}, 4 -- \citet{2004ApJ...609..564S}}

\end{table}

\begin{table}[t]
\caption{
The list of 37 experiments in which the high-redshift radio sources from Table~\ref{table_sources} were observed. Telescope codes listed in the column ``Array'' are explained in Table~\ref{table_stations}.
The maximum baseline length is shown in the last column.
}
\vspace{3mm}
\begin{tabular}{l l c l c}
\hline
Session	   &  Date   & Array	& High-$z$ sources (on-source time, h)  & Maximum baseline (km) \\
\hline 
\hline
UF001A & 2017 Jan 16 & VLBA & 0901$+$697 (0.08) & 8611 \\
RUA011 & 2017 Jan 28 & Sv-Zc-Bd & 0901$+$697 (3.3) &  4405 \\
RUA012 & 2017 Feb 04 & Sv-Zc-Bd-Ys & 0901$+$697 (2.48), 2101$+$600 (2.87) & 7080 \\
RUA015 & 2017 Apr 08 & Sv-Zc-Bd-Ys-Sh & 0901$+$697 (3.89), 2101$+$600 (5.06) & 9170 \\
RUA016 & 2017 Apr 22 & Sv-Zc-Bd-Ys-Sh & 0901$+$697 (3.28), 2101$+$600 (4.08) &  9170 \\
UF001G & 2017 Apr 28 & VLBA & 1508$+$572 (0.02) &  8611 \\
UF001H & 2017 May 01 & VLBA & 1508$+$572 (0.02) &  8611 \\
UF001O & 2017 Aug 05 & VLBA & 1508$+$572 (0.02) &  8611 \\
RUA017 & 2017 Aug 19 & Sv-Zc-Bd-Ys-Sh & 0901$+$697 (3.1), 1428$+$422 (1.12), 2101$+$600 (4.08) & 9170 \\
RUA018 & 2017 Aug 26 & Sv-Bd-Ys-Sh & 0901$+$697 (4.84), 1428$+$422 (1.57), 2101$+$600 (4.67) &  9170 \\  
RUA019 & 2017 Nov 18 & Sv-Bd-Ys-Sh & 0901$+$697 (5.23), 1428$+$422 (0.07), 2101$+$600 (4.86) &  9170 \\  
RUA020 & 2017 Nov 25 & Sv-Zc-Bd-Ys & 0901$+$697 (5.04), 1428$+$422 (0.07), 2101$+$600 (4.67) & 7080 \\  
UG002B & 2018 Jan 25 & VLBA & 0901$+$697 (0.07), 1508$+$572 (0.03) &  8611 \\ 
RUA022 & 2018 Feb 03 & Sv-Zc-Bd-Ys-Sh & 0901$+$697 (3.88), 2101$+$600 (5.05) & 9170 \\ 
RUA023 & 2018 Mar 31 & Sv-Zc-Bd-Ys-Sh & 0901$+$697 (3.68), 2101$+$600 (5.59) & 9170 \\
UG002H & 2018 May 19 & VLBA & 1508$+$572 (0.02) & 8611 \\
UF002I & 2018 Jun 03 & VLBA & 1508$+$572 (0.02) & 8611 \\
ET036A & 2018 Jun 04 & Wn-Mc-Nt-On-Ef- & 0901$+$697 (3.75), 1428$+$422 (3.48), \\
 & & Sv-Zc-Bd-Ys-Hh & 2101$+$600 (3.87) &  7080 \\
RUA024 & 2018 Jun 23 & Sv-Zc-Bd-Ys-Sh & 0901$+$697 (4.26), 2101$+$600 (7.18) &  9170 \\
V515C & 2018 Jul 19 & Sv-Zc-Bd-Sh-Km- & 0901$+$697 (1.25),  \\
 & & Ho-Hh &  2101$+$600 (2.0)  & 6711 \\
RUA025 & 2018 Jul 21 & Sv-Zc-Bd-Ys & 0901$+$697 (3.68), 2101$+$600 (5.38) &  7080 \\
RUA026 & 2018 Sep 08 & Sv-Zc-Bd-Ys & 0901$+$697 (3.88), 2101$+$600 (5.39) & 7080 \\
ET036B & 2018 Oct 27 & Wz-Mc-On-Ef-Sv- & 0901$+$697 (3.73), 1428$+$422 (1.45),  \\  
 & & Zc-Bd-Ys-Hh-T6 & 2101$+$600 (4.28) &  9170 \\
UG002T & 2018 Nov 18 & VLBA & 1508$+$572 (0.02) & 8611 \\
RUA028 & 2019 Jan 05 & Sv-Zc-Bd-Ys & 0901$+$697 (4.07), 2101$+$600 (4.49) & 7080 \\
ET036C & 2019 Feb 25 & Wn-Mc-On-Ef-Sv- & 0901$+$697 (3.38), 1428$+$422 (1.65),  \\
 & & Zc-Bd-Ys-Hh-T6-Ur & 2101$+$600 (3.62) &  9170 \\
RUA030 & 2019 Jun 29 & Sv-Zc-Bd-Ys-Sh & 0901$+$697 (3.29), 1508$+$572 (2.21), 2101$+$600 (5.69) &  9170 \\
RUA031 & 2019 Jul 06 & Sv-Zc-Bd-Km-Ho & 0901$+$697 (2.52), 1508$+$572 (2.08), \\
 & & & 2101$+$600 (3.89) &  6761 \\
RUA032 & 2019 Jul 13 & Sv-Zc-Bd-Ys-Sh-Hh & 0901$+$697 (1.94), \\
 & & & 1508$+$572 (2.34) &  9170 \\
RUA033 & 2019 Sep 28 & Sv-Zc-Bd-Ys-Sh & 0901$+$697 (2.91), 1508$+$572 (2.62), 2101$+$600 (4.79) & 9170 \\
RUA037 & 2020 Jan 25 & Sv-Zc-Bd-Ys & 0901$+$697 (3.26), 1508$+$572 (2.21), 2101$+$600 (5.03) &  7080 \\
RUA038 & 2020 Jun 27 & Sv-Zc-Bd-Ys-Sh & 0901$+$697 (4.56), 1508$+$572 (1.88), 2101$+$600 (5.04) &  9170 \\
RUA039 & 2020 Nov 28 & Sv-Zc-Bd-Ys & 0901$+$697 (2.13), 1508$+$572 (2.21), 2101$+$600 (3.29) & 7080 \\
RUA040 & 2020 Dec 26 & Sv-Zc-Bd-Ys-Sh & 0901$+$697 (2.71), 1508$+$572 (1.77), 2101$+$600 (5.39) & 9170 \\
RUA041 & 2021 Mar 20 & Sv-Zc-Bd-Sh & 0901$+$697 (2.99), 1508$+$572 (0.69), 2101$+$600 (3.97) & 6761 \\
RUA042 & 2021 Mar 27 & Sv-Zc-Bd-Sh & 0901$+$697 (1.34), 1508$+$572 (0.24), 2101$+$600 (2.59) & 6761 \\
RUA043 & 2021 Apr 10 & Sv-Zc-Bd-Sh & 0901$+$697 (0.74), 2101$+$600 (1.37) & 6761 \\
\hline

\label{table_sessions}
\end{tabular}
\end{table}  

\begin{table}[h]
\caption{Radio telescopes involved in the study presented in this paper, in the alphabetical order of their codes used in Table~\ref{table_sessions}. 
}
\begin{tabular}{llcll} 
\hline
Telescope	& Telescope & Telescope      & Institute or observatory & Country  \\
code        & name      & diameter [m]   &             &          \\
\hline
Bd & Badary & 32 & Institute of Applied Astronomy & Russia \\ 
Ef & Effelsberg & 100 & Max Planck Institute for Radio Astronomy & Germany \\
Hh & Hartebeesthoek & 26 & South African Radio Astronomy Observatory & South Africa \\
Ho & Hobart & 26 & Mt. Pleasant Radio Observatory, Tasmania & Australia \\
Km & Kunming & 40 & Yunnan Astronomical Observatory & China \\
Mc & Medicina & 32 & INAF Institute of Radio Astronomy & Italy \\ 
Nt & Noto & 32 & INAF Institute of Radio Astronomy & Italy \\
On &  Onsala & 20 & Onsala Space Observatory & Sweden \\
Sh & Sheshan & 25 & Shanghai Astronomical Observatory & China \\
Sv & Svetloe & 32 &Institute of Applied Astronomy & Russia \\
T6 & Tianma & 65 & Shanghai Astronomical Observatory & China \\
Ur & Nanshan (Urumqi) & 25 & Xinjiang Astronomical Observatory & China \\ 
VLBA & VLBA & $10\times 25$ & National Radio Astronomy Observatory & USA \\
Wn & Wettzel-North & 13 & Geodetic Observatory Wettzell & Germany \\  
Wz & Wettzel & 20 & Geodetic Observatory Wettzell & Germany \\ 
Ys & Yebes & 40 & Yebes Observatory & Spain \\
Zc & Zelenchukskaya & 32 & Institute of Applied Astronomy & Russia \\
\hline 
\label{table_stations}
\end{tabular}
\end{table}

Table~\ref{table_sessions} lists the observing sessions, their dates, radio telescopes involved, observed target sources and on-target times. Table~\ref{table_stations} explains the codes of radio telescope shown in Table~\ref{table_sessions}.
RUA project code is associated with the regular 24~h astrometric sessions by the network of five radio telescopes, including Yebes (Spain), Sheshan (China), Badary, Svetloe, and Zelenchukskaya (Russia). The latter three radio telescopes are the part of the Russian National Quasar VLBI Network \citep{SHUYGINA2019150}. The Hobart (Australia) and Kunming (China) radio telescopes participated in RUA031, the Hartebeesthoek (South Africa) radio telescope participated in RUA032. 
The experiment V515C was conducted using the Australian Long Baseline Array (LBA) -- a part of the Australia Telescope National Facility. Three ET036 experiments were carried out by the European VLBI Network (EVN).
We also used data from several observing sessions of the Very Long Baseline Array (VLBA) conducted by the United States Naval Observatory (USNO) and the National Aeronautics and Space Administration Goddard Space Flight Center (NASA GSFC) in the projects not related to ours.

\begin{table}[h]
\caption{Recording setups of the sessions}
\begin{tabular}{lcccc} 
\hline
Session  &  Sample rate    & Sampling  &  3.5\,cm   &  13\,cm   \\
names    &  Mbit\,s$^{-1}$  & bits       &  setup     &  setup    \\
\hline
RUA011          &  256   & 1          &  $10 \times 8$~MHz  &  $6 \times 8$~MHz  \\
ET036A--ET036C  &  512   & 2          &  $10 \times 8$~MHz  &  $6 \times 8$~MHz  \\
RUA012--RUA041  &  1024  & 2          &  $10 \times 16$~MHz &  $6 \times 16$~MHz  \\
V515C           &  1024  & 2          &  $10 \times 16$~MHz &  $6 \times 16$~MHz  \\
RUA042          &  1024  & 2          &  $16 \times 16$~MHz &  ---   \\
RUA043          &  2048  & 2          &  $16 \times 32$~MHz &  ---   \\
UF001A--UG002T  &  2048  & 2          &  $12 \times 32$~MHz &  $4 \times 32$~MHz  \\
\hline 
\label{table_setups}
\end{tabular}
\end{table} 

RUA and EVN sessions listed in Table~\ref{table_sessions} were scheduled with the \textsc{sked} software \citep{2010ivs..conf...77G} in the Institute of Applied Astronomy of the Russian Academy of Sciences (IAA RAS) in Saint-Petersburg. The LBA session (V515C) was scheduled using \textsc{sked} in the Shanghai Astronomical Observatory of the Chinese Academy of Sciences (SHAO CAS). Standard geodetic/astrometric approach of scheduling the numerous scans on the target radio sources mixed with calibrator source scans was used. All RUA and EVN sessions used standard IVS 3.5/13\,cm wide-band setup in right circular polarization, except for two experimental RUA sessions that used only 3.5\,cm wavelength. The LBA session used modified 3.5/13\,cm wide-band setup in right circular polarization.
The standard IVS 3.5/13\,cm setup includes 8 upper side band and 2 lower side band frequency channels (IFs) at 3.5\,cm and 6 upper side band IFs at 13\,cm, specifically placed over bandwidths of 720~MHz and 140~MHz, respectively. The 13\,cm frequency coverage of LBA session is reduced to 90~MHz due to radio-frequency interference (RFI).
Recording setups of the RUA, EVN, LBA, and VLBA sessions are listed in Table~\ref{table_setups}. Sessions differ by the IF bandwidths, bit sampling, and total sample rate. The RUA042 session used 3.5\,cm frequencies of IVS geodetic wide-band setup and included 8 upper side band and 8 lower side band IFs distributed over 720~MHz bandwidth. The RUA043 session used 32-MHz bandwidth IFs forming almost continuous frequency coverage between 8188 and 8972~MHz.
The detailed description of the VLBA experiments can be found in \citet{2021AJ....162..121H}. 

Three of the four observed radio sources (the top three in Table~\ref{table_sources})  have multiple spectroscopic measurements of their redshifts with reasonable confidence. Recently the Sloan Digital Sky Survey \citep[SDSS,][]{2000AJ....120.1579Y} clearly displayed the dominating Ly-$\alpha$ peak and other characteristic properties of high-redshift objects, e.g. Ly-break at $\lambda = 912$~\AA\ rest-frame wavelength, Ly-$\beta$ and C{IV} emission lines. The only known spectrum of the fourth object, 2101$+$600, has unreliable identification of the Ly-$\alpha$ emission line \citep{2004ApJ...609..564S}.
However, the Panoramic Survey Telescope and Rapid Response System \citep[Pan-STARRS]{2016arXiv161205560C} survey provides photometric magnitudes for 2101$+$600 as $g = 24.9$ and $r = 22.6$. The large colour difference $g-r = 2.3$ might hint on the drop in continuum between $g$ and $r$, supporting the high redshift tentatively determined by \citet{2004ApJ...609..564S}.

The redshifts for the three objects (0901$+$697, 1428$+$422 and 1508$+$572) are compromised by
damping of the Ly-$\alpha$ line, the blueshift effect of C{IV}, and low intensity in Ly-$\beta$. Therefore, a conservative estimate of the redshift formal errors at $z\sim~3$ is about $\sigma_{z}=0.005$  \citet[e.g.][]{2013AJ....146...10T,2017AJ....153..157T}.

\section{Data Processing and Analysis}

The data of RUA sessions were transferred to and correlated at the Correlator Processing Center of the Russian Academy of Sciences at the IAA RAS in Saint-Petersburg. Correlation was performed using ARC \citep{2008mefu.conf..124F} and DiFX \citep{2011PASP..123..275D} correlators. 
Session RUA011 was correlated using ARC, DiFX version 2.4.0 was used for sessions RUA012~---~RUA025, version 2.5.2 for RUA026~---~RUA033, version 2.6.2 for RUA037~---~RUA043, respectively. Three EVN sessions of the ET036 project were correlated using the EVN Software Correlator at JIVE \citep[SFXC,][]{2015ExA....39..259K}. The LBA session was correlated at the Pawsey Supercomputing Centre with DiFX. 

Visibility data of 0.5~s integration time and of 125~kHz spectral resolution correlated using DiFX were post-processed with the \textsc{pima} software \citep{2011AJ....142...35P} following the Data analysis pipeline from the {\sc PIMA} User Guide\footnote{\url{http://astrogeo.org/pima/pima_user_guide.html}, accessed on 2022.10.09.}.
The LBA and EVN output data in geodetic Mark4 format were post-processed with the Haystack Observatory Postprocessing System\footnote{\url{https://www.haystack.mit.edu/haystack-observatory-postprocessing-system-hops/}, accessed on 2022.10.09.} \citep[HOPS,][]{2004RaSc...39.1007W}.

Standard approach of calibrating the ionospheric delay was used for the RUA observations at 3.5/13\,cm wavelength \citep{1996jpl..rept.8339S} (sessions RUA011~---~RUA030). The ionospheric contribution was calculated using daily Global Ionospheric Maps (GIMs) produced by the Center for Orbit Determination in Europe (CODE) Analysis Center \citep{Schaer1996} for the rest of observations (sessions RUA031~---~RUA043). 
Various research have shown that ionospheric contribution calculated using GIMs and derived from geodetic VLBI are in good agreement. In particular, \citet{Hobiger_2006} analysed available VLBI sessions from 1995 to 2006, their estimate of mean difference between the two techniques is less than 3 TECU ($\sim$57\,ps). Recently, \citet{Gordon_IVS_GM2010} analysed ten 3.5/13\,cm IVS RDV (Research and Development) sessions and found a moderate increase of the source position uncertainties at 3.5\,cm when using GIMs by $\sim$0.5\,mas. However, his analysis of the thirteen 3.5/13\,cm VLBA Calibrator Survey (VCS) sessions did not show a significant difference between the standard dual frequency calibration of the ionosphere and the newly developed GIMs approach. Moreover, several radio sources were detected at 3.5\,cm only.

In some RUA experiments, 13\,cm band data were seriously corrupted by RFI, making it impossible to get a solution from 3.5/13\,cm data. We decided to work only with 3.5\,cm data using the algorithm described in \citet{Aksim_2019}. 
Taking possible contamination by RFI into account, some of the latest RUA experiments were specially designed as a single 3.5\,cm band experiment. That allows us to increase the number of observations of faint target sources due to higher recording data rate and therefore shorter on-source time.
We used GIMs to calculate ionospheric contribution for the experiments from 2019.5 to 2021.5. This period corresponds to a low solar activity, in contrary to the above mentioned works where the VLBI data covered almost a full solar cycle of activity. We expect that the ionospheric contribution using GIMs during the period of low solar activity is negligible. To ensure this, we compared source position estimates for a number of RUA experiments in this period and got an agreement within 1-$\sigma$ formal errors.
The total delays obtained from 3.5\,cm data along with the ionospheric delay from either dual-band combination or using vertical total electron content (TEC) from GIMs were then astrometrically analysed.

The data of all presented in this paper VLBI experiments were processed with the \textsc{OCCAM} 6.3 software  \citep{2004ivsg.conf..267T} in the session-wise mode. The astronomical reduction of the data (precession, nutation, tidal correction, etc.) was realized in accordance with the IERS Conventions~2010 \citep{iers10}. The astrometric and geodetic parameters were estimated by the least squares collocation method \citep{2000ITN....28...33T}. The list of parameters comprises the Earth pole position components, correction to the Universal Time (UT1--UTC), nutation offsets, and daily positions of the radio telescopes. All the parameters were estimated for each individual experiment along with the position corrections to the high redshift radio sources from Table~\ref{table_sources}. A priori positions of the radio telescopes were referred to the International Terrestrial Reference Frame \citep[ITRF2014,][]{2016JGRB..121.6109A}. We estimated the positions of the radio sources from Table~\ref{table_sources} only, the positions of all other observed radio sources were fixed to the aus2020a catalogue values\footnote{\label{fnote1}\url{https://cddis.nasa.gov/archive/vlbi/ivsproducts/crf/}}.
This catalogue comprises new observations during 2018--2020, after the official release of ICRF3. The most essential change with respect to ICRF3 is an improvement of positions of rarely observed radio sources in the southern hemisphere, while positions of the defining radio sources in the northern hemisphere are very close to the original ICRF3 catalogue values. Among $\sim 150$ defining radio sources from the northern hemisphere, the positions of more than 100 in the aus2020a catalogue do not differ from the ICRF3 positions by more than $0.025$~mas in both coordinates, so one can state that the daily positions of the four high-redshift objects studied in this paper are very close to the ICRF3 system.

A priori dry zenith delays were determined from local atmospheric pressure values \citep{1972GMS....15..247S}, which were then mapped to the elevation of the observed sources using the Vienna mapping function \citep[VMF1,][]{2006JGRB..111.2406B}. The elevation cut-off angle was set to $5\degr$. To calibrate the troposphere instability induced by the non-equilibrium water vapour during a 24-h experiment, we estimated the so-called zenith wet delays with North--South and East--West gradients at each observational epoch \citep{2000ITN....28...33T}. The instability of the hydrogen maser clock at all VLBI stations was also estimated for each observational epoch.

Section~\ref{sec:results} provides detailed results for individual sources, along with an interpretation in the astrophysical context.
We approximated the time series of daily right ascension and declination components of the four radio sources with linear functions to calculate the apparent proper motion by the conventional weighted least squares method. Variations of $2101+600$ were also treated in a special way (see below).

\section{Observational Results}
\label{sec:results}

\subsection{0901$+$697}

This is the most distant quasar \citep[$z = 5.47$,][]{2004ApJ...610L...9R} among those available for geodetic VLBI observations, due to its comparatively high total flux density. We observed 0901$+$697 in 23 RUA experiments, three EVN experiments, and added two VLBA experiments conducted in independent projects (UF001A and UG002B). The VLBA data are publicly available from the U.S. National Radio Astronomy Observatory archive\footnote{\url{https://data.nrao.edu/}} and the IVS database.

The source has been imaged with VLBI at various frequencies, up to 43\,GHz and multiple epochs since 2004 \citep{2004ApJ...610L...9R,2017MNRAS.468...69Z,2018A&A...618A..68F,2020NatCo..11..143A}. Its pc-scale structure is dominated by a prominent compact core. The jet is extended to southwest with a component separated by $\sim 0.7$\,mas, then apparently bends southwards, ending in a component located within $\sim 1.5$\,mas from the core. By detecting polarized radio emission on pc scales and estimating jet component proper motions for the first time in this source, \citet{2020NatCo..11..143A} found that the properties of 0901$+$697 are the best explained with a nascent jet embedded in and interacting with a dense surrounding matter. Note that we use the term ``jet component proper motion'' when speaking about an apparent change in the relative separation of core and jet components in VLBI images measured between different observing epochs. This is not to be confused with the apparent proper motion of the quasar as a whole, like the values given in Table~\ref{table_pm}.

While the position of the inner jet component, possibly marking the location of a standing shock produced by the jet encountering the ambient medium, is nearly stationary, the more distant southern jet component has a small but significant proper motion, $0.019 \pm 0.006$\,mas\,yr$^{-1}$, measured between 2004 and 2018. According to the 2.3-GHz image sensitive to more extended steep-spectrum radio emission \citep{2018A&A...618A..68F}, the jet bending continues towards the southeast, up to $\sim 2.5$\,mas from the core. According to occasional multi-epoch VLBI observations at various radio frequencies from 8.4 to 43~GHz, the source flux density is changing on time scales of years \citep[see the data in][]{2017MNRAS.468...69Z,2020NatCo..11..143A}. But the total 15-GHz flux density regularly monitored at the 40-m antenna of the Owens Valley Radio Observatory \citep[OVRO,][]{2011ApJS..194...29R}, which is dominated by the compact flat-spectrum core at this high frequency, only slightly decreased in 2017--2018, the period overlapping with our astrometric observations \citep[see the supplementary figure 1 in][]{2020NatCo..11..143A}. The comparison of OVRO single-dish and VLBI flux densities obtained with very different angular resolutions is justified by the fact that near-simultaneous 15-GHz measurements of 0901+697 indicate no appreciable difference between them \citep{2020NatCo..11..143A}, suggesting negligible contribution of any emission extended to spatial scales beyond those probed by VLBI. This applies for strongly core-dominated variable radio sources in general \citep[e.g.][]{2011ApJ...742...27L}.

No trend was found in any of the coordinates after linear least squares fitting (Table~\ref{table_pm}). The potential impact of the source structure (the core fading and the outward motion of the southern jet component) needs to be verified with future observations.

\begin{figure}
\centering
 \includegraphics[width=0.4\textwidth, clip=, angle=0]{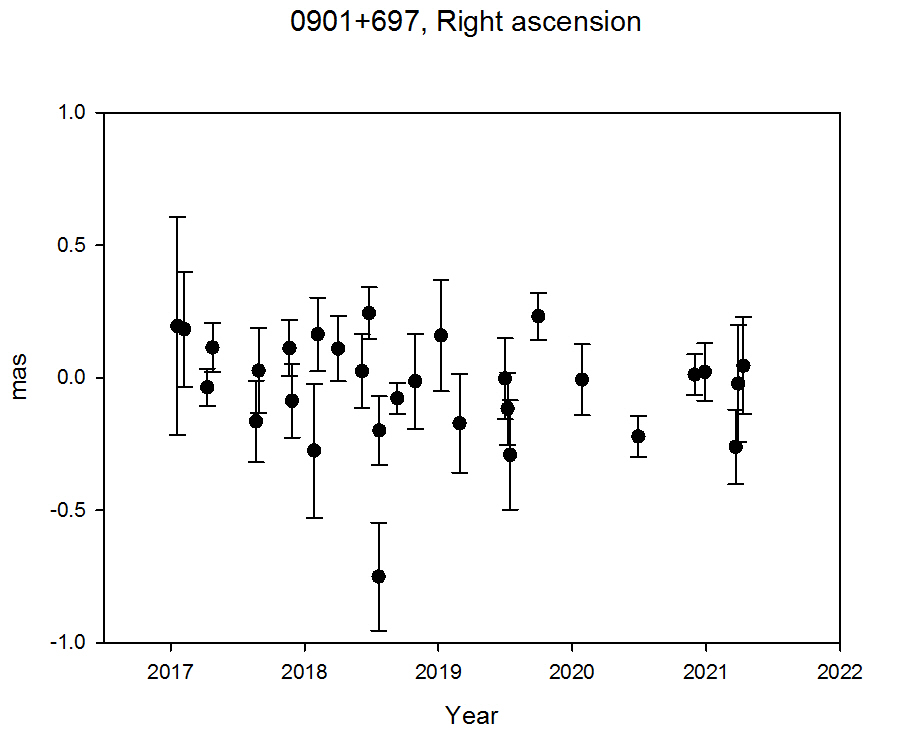}
 \includegraphics[width=0.4\textwidth, clip=, angle=0]{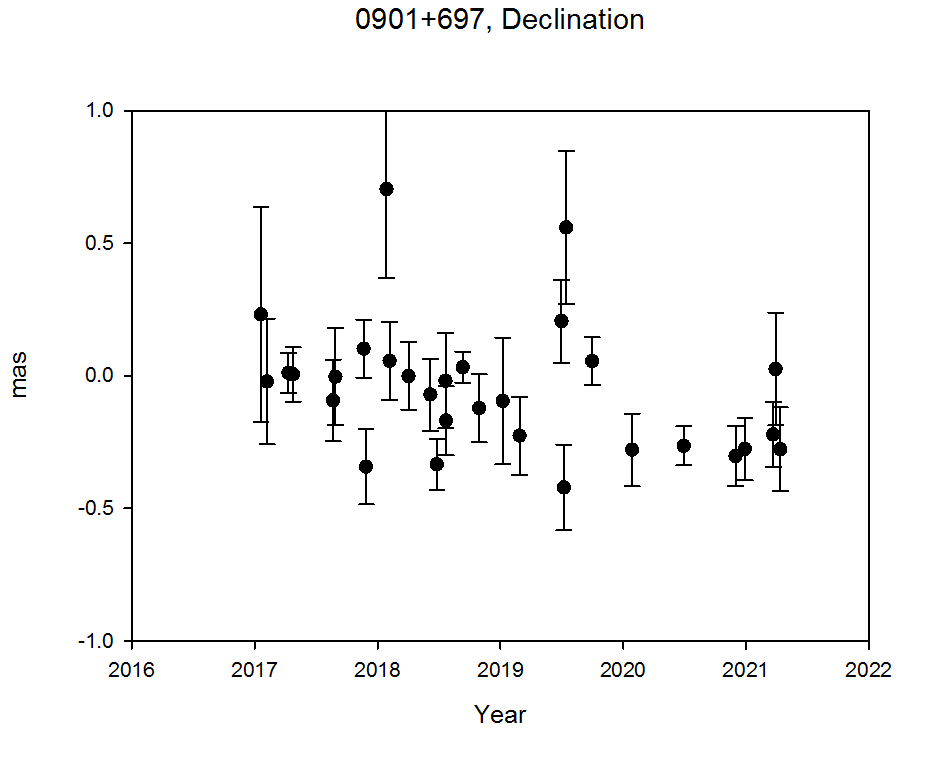}
 \caption{Relative astrometric coordinates (right ascension and declination) of 0901$+$697 as a function of time, between 2017 and 2020. The reference point corresponds to the solution aus2020a.crf. The error bars show $\pm 1\sigma$ formal uncertainty.} 
 \label{fig:0901}
\end{figure}

\subsection{1428+422}

\begin{figure}
\centering
 \includegraphics[width=0.4\textwidth, clip=, angle=0]{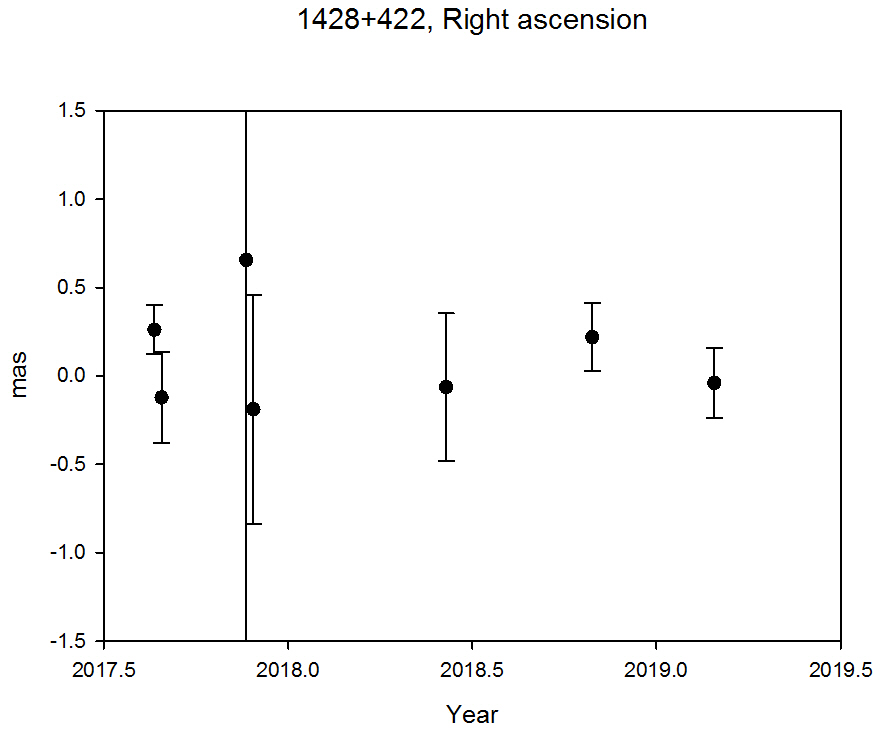}
 \includegraphics[width=0.4\textwidth, clip=, angle=0]{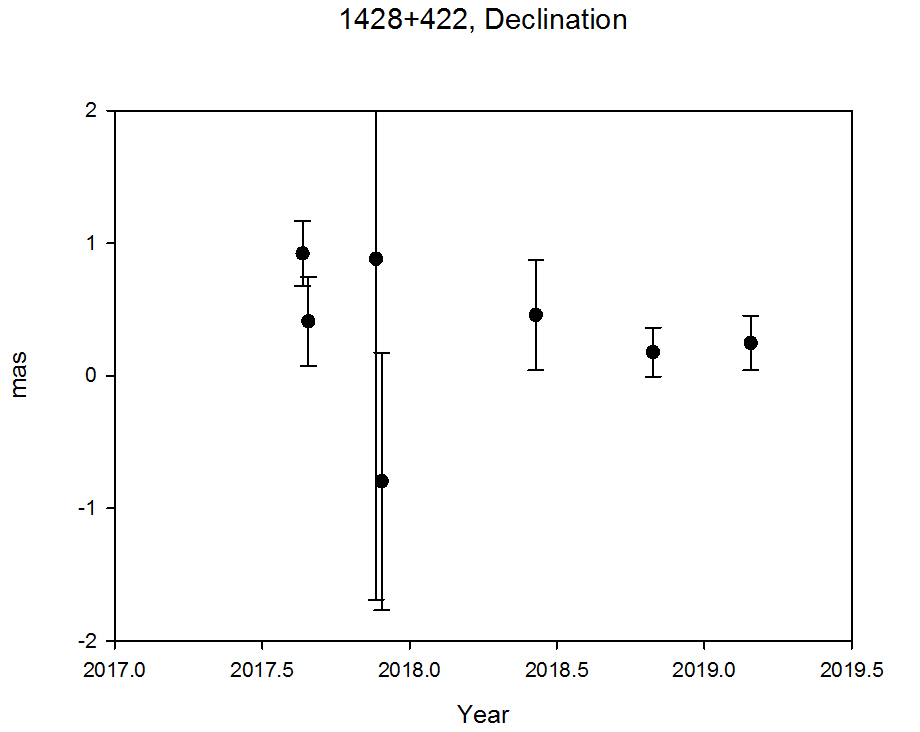}
 \caption{Time series of the relative right ascension (left) and declination (right) of 1428$+$422. The reference point corresponds to the solution aus2020a.crf. The error bars show $\pm 1\sigma$ formal uncertainty.}
 \label{fig:1428}
\end{figure}

The source 1428$+$422 was identified as a high-redshift (z$=$4.715) quasar by \citet{1998MNRAS.294L...7H}. 
It has been observed with VLBI as a part of the astrometric program occasionally since 1996, but the number of delays in the majority of experiments did not exceed 10, therefore these early data are not included in the current analysis. We collected 7 data points (4 from RUA experiments and 3 from EVN) between 2017 and 2019. No significant proper motion was detected from this set of experiments (Fig.~\ref{fig:1428}). 

The most sensitive 2.3-GHz VLBA image of the source shows a core--jet structure extending to $\sim 20$\,mas in the southwest direction \citep{2020SciBu..65..525Z}. The quasar displays remarkable variations in its flux density producing a major outburst in 2005. The 15-GHz VLBA imaging observations performed shortly before and after the outburst did not reveal a birth of a new component in the inner jet \citep{2010A&A...521A...6V}. However, based on  
5 epochs of 8.4~GHz VLBI imaging data spanning $22$\,yr, \citet{2020SciBu..65..525Z} studied the jet kinematics and found two components within $\sim 3$\,mas separation from the core moving with apparent jet component proper motions $0.017 \pm 0.002$\,mas\,yr$^{-1}$ and $0.156 \pm 0.015$\,mas\,yr$^{-1}$, with the quickly fading outer jet component being the fastest. The latter value corresponds to an apparent transverse speed of $19.5 \pm 1.9$ times the speed of light, marking the jet in 1428+422 one the fastest-moving jets in $z>4$ sources among known to date \citep{2020SciBu..65..525Z}.

\subsection{1508$+$572}

\begin{figure}
\centering
 \includegraphics[width=0.8\textwidth, bb=0 0 700 700, clip=, angle=0]{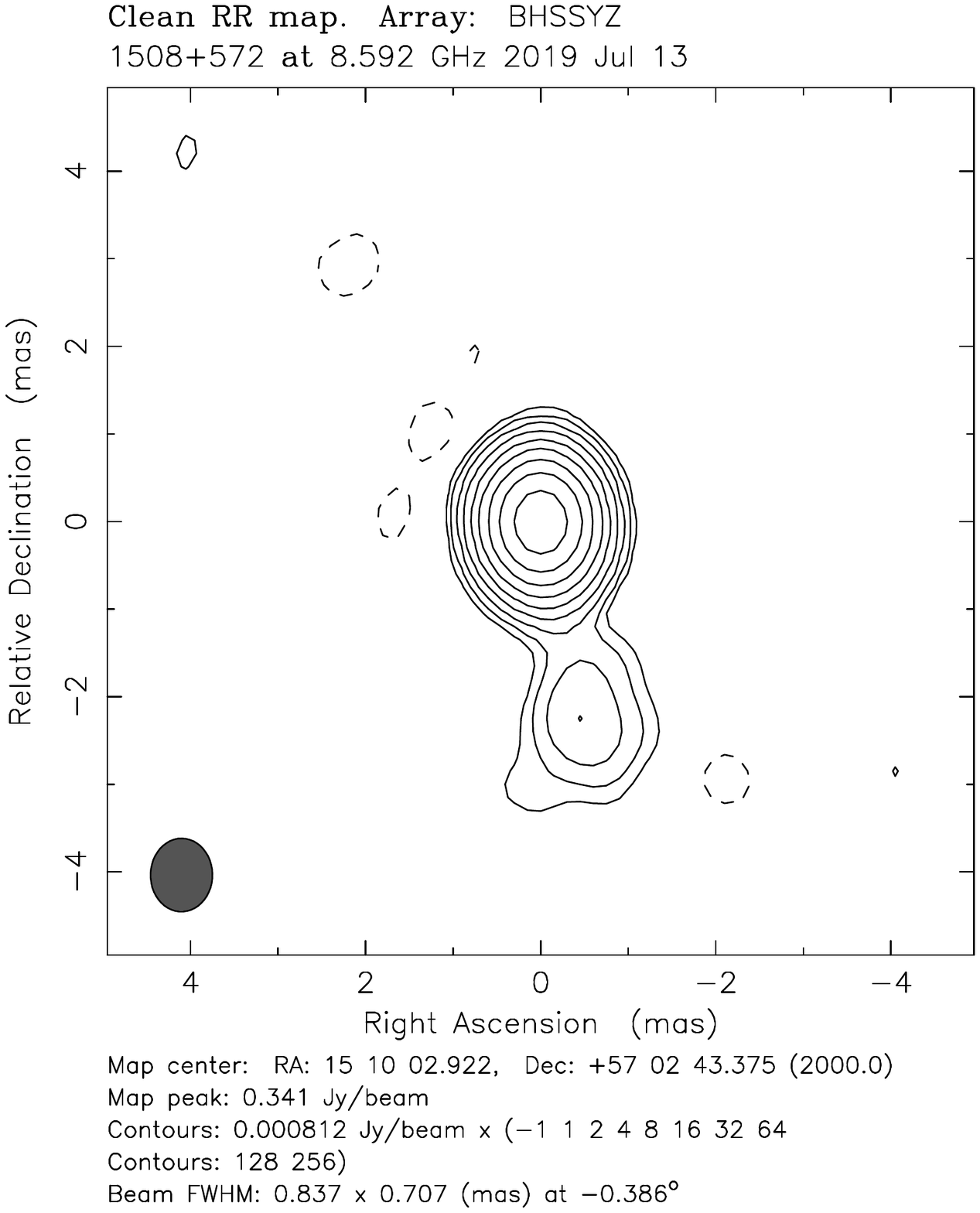}
 \caption{A 8.6-GHz VLBI image of 1508$+$572 obtained with a 5-element global network of radio telescopes (Sv, Zc, Bd, Yb, Sh) on 2019 Jul 13 (project code RUA032). The peak brightness is $341$\,mJy\,beam$^{-1}$, the lowest contours are at $0.812$\,mJy\,beam$^{-1}$, the positive contour levels increase by a factor of 2. The elliptical Gaussian restoring beam indicated in the lower left corner has major and minor axes $0.837$\,mas and $0.707$\,mas (half-power beam width), respectively, with a major axis position angle $-0\fdg4$, measured from north through east. The mas-scale structure of the source is dominated by the bright core, a jet component is seen towards south.} 
\label{1508-image}
\end{figure}

The source 1508$+$572 was identified as a radio-loud quasar at $z=4.30$ by \citet{1995MNRAS.273L..63H}. Due to an insufficient angular resolution, its first 5-GHz VLBI EVN image  could only provide a hint on the core--jet structre pointing to south \citep{1997A&A...325..511F}. Since then, the compact mas-scale structure of 1508$+$572 extending to $\sim 3$\,mas has been imaged with VLBI at multiple frequencies at several epochs \citep[e.g.][]{2002ApJS..141...13B,2005AJ....129.1163P,2007ApJ...658..203H,2011MNRAS.415.3049O,2012A&A...544A..34P,2016AJ....151..154G}. As an example, the 8.6-GHz image displayed in  Fig.~\ref{1508-image} was made from the RUA032 experiment data obtained on 2019 Jul 13. The Astronomical Image Processing System \citep[\textsc{AIPS,}][]{2003ASSL..285..109G} software package was used for calibrating the visibility data in a standard way, and hybrid mapping was performed with the \textsc{Difmap} program \citep{1997ASPC..125...77S}. We followed the same data analysis procedure as described in \citet{2018A&A...618A..68F}.

The source 1508$+$572 was monitored frequently with astrometric VLBI in the 1990s but, similarly to the case of 1428$+$422, the number of experiments and delays dropped dramatically after 2000. However, since 2017, 1508$+$572 has been observed in a set of astrometric VLBI experiments \citep{2021AJ....162..121H}. We have started our own observations of this radio source only from the middle of 2019. 

\begin{figure}
\centering
 \includegraphics[width=0.4\textwidth, clip=, angle=0]{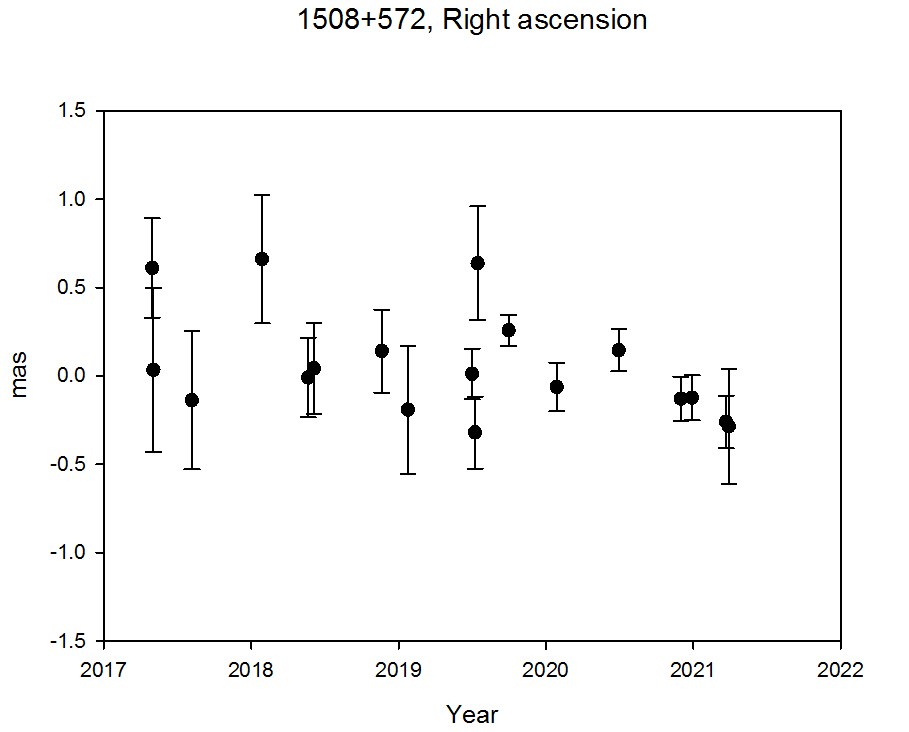}
 \includegraphics[width=0.4\textwidth, clip=, angle=0]{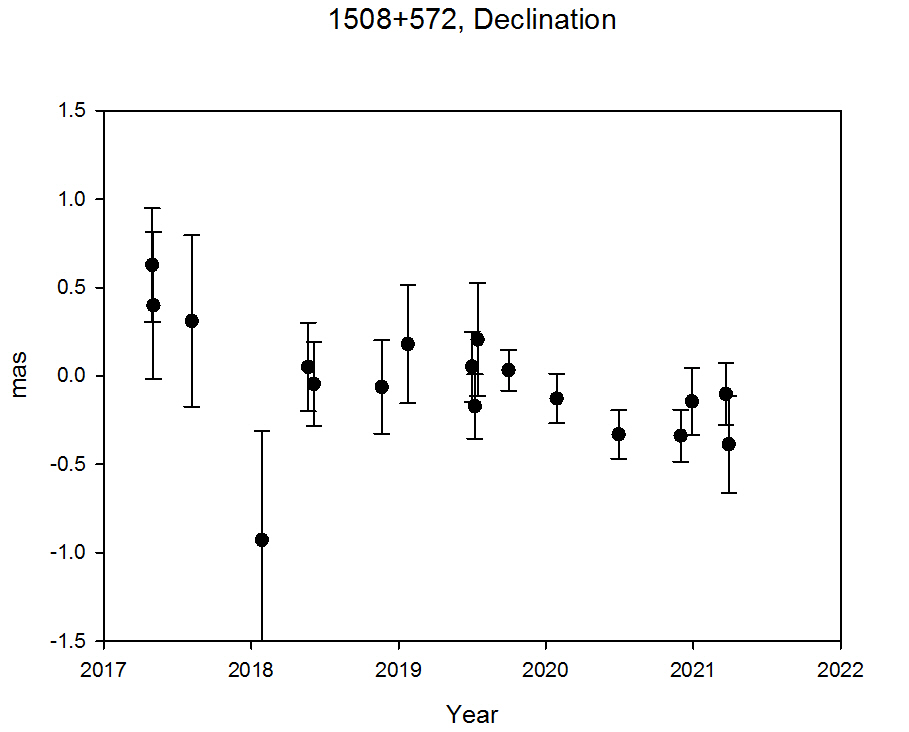}
 \caption{Relative astrometric coordinates of 1508$+$572 as a function of time, between 2017 and 2020. The reference point corresponds to the solution aus2020a.crf. The error bars show $\pm 1\sigma$ formal uncertainty.}  
\label{fig:1508}
\end{figure}

The coordinate time series shown in Fig.~\ref{fig:1508} displays a steady linear trend with rates $-0.121 \pm 0.051 $\,mas\,yr$^{-1}$ in right ascension and $-0.147 \pm 0.040$\,mas\,yr$^{-1}$ in declination. The latter estimate is statistically significant (i.e. $>3\sigma$) and it may be related to the jet structure positioned along the north--south direction.

\begin{figure}
\centering
 \includegraphics[width=0.6\textwidth, clip=, angle=0]{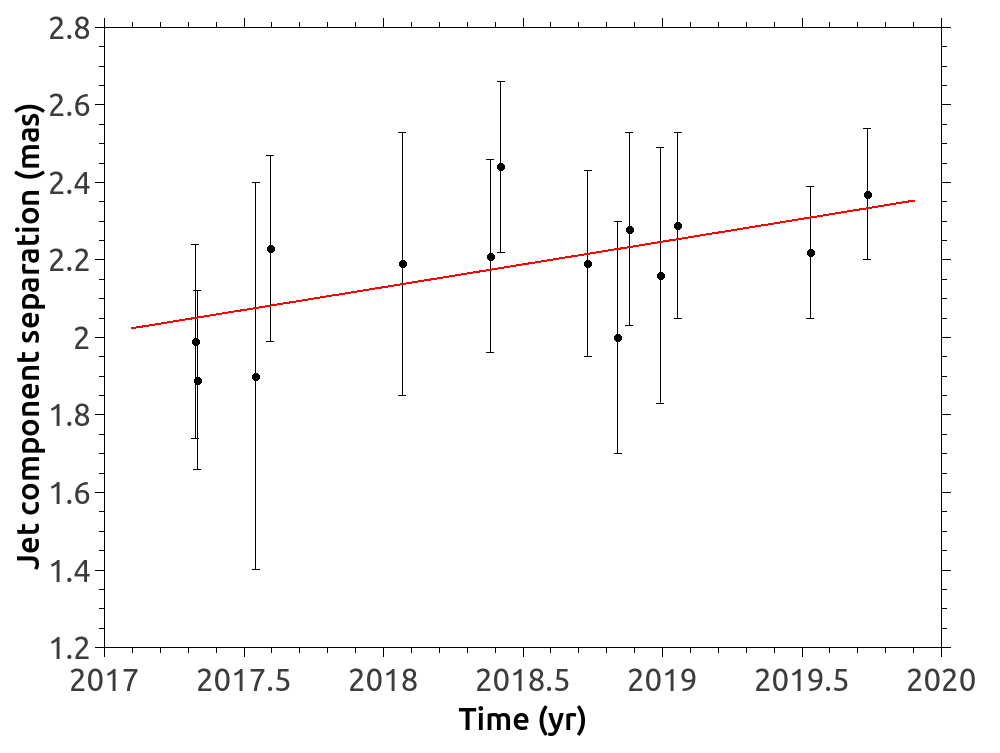}
 \caption{Angular separation of the fitted core and jet components in 1508$+$572 at 8.6/8.7~GHz as a function of time. The line indicates the linear least squares fit with the slope $0.117 \pm 0.078$\,mas\,yr$^{-1}$.}
\label{1508:separation}
\end{figure}

\begin{figure}
\centering
 \includegraphics[width=0.6\textwidth, clip=, angle=0]{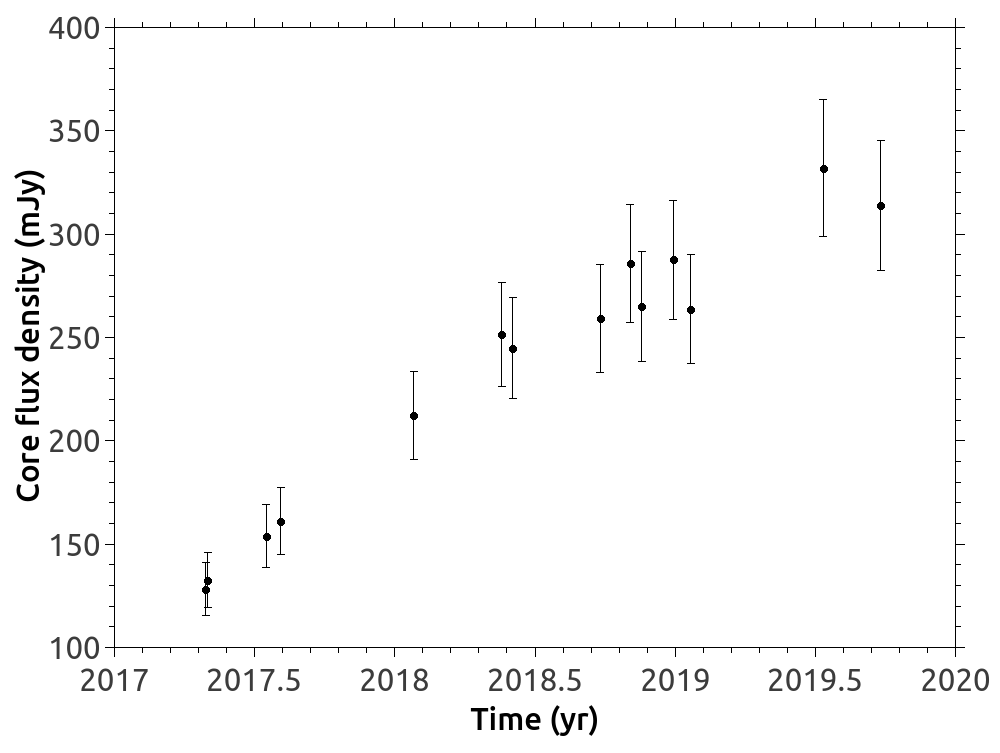}
 \caption{Fitted core component flux density in 1508$+$572 at 8.6/8.7~GHz as a function of time.} 
\label{1508:lightcurve}
\end{figure}

Since jet proper motion studies are not yet available for 1508$+$572 in the literature, we downloaded calibrated $X$-band (8.6/8.7~GHz) VLBI visibility data from the Astrogeo archive\footnote{\url{http://astrogeo.org/cgi-bin/imdb_get_source.csh?source_name=1508\%2B572}, maintained by Leonid Petrov} that overlap in time with our astrometric study (Table~\ref{table_sessions}). 
These data from VLBA experiments UF001G (2017 Apr 29), UF001H (2017 May 2), UF001N (2017 Jul 17), UF001O (2017 Aug 5) \citep{2021AJ....162..121H}, UG002B (2018 Jan 25), UG002H (2018 May 20), UG002I (2018 Jun 3), UG002R (2018 Sep 25), UG002S (2018 Nov 3), UG002T (2018 Nov 18), UG002W (2018 Dec 29), and UG002X (2019 Jan 21) were supplemented with data from RUA experiments where we obtained sufficient interferometric visibility points for good-quality imaging, RUA032 (2019 Jul 13, Fig.~\ref{1508-image}) and RUA033 (2019 Sep 28). 
Imaging was done in \textsc{Difmap} \citep{1997ASPC..125...77S}, where the brightness distribution of 1508$+$572 was also modeled with elliptical and circular Gaussian components for the core and the southern jet feature, respectively. Model fits allow us to quantitatively characterise how the source changes with time. For error estimates, we followed \citet{2021ApJ...919...40P}, and took $1/5$ of the projection of the elliptical Gaussian synthesized beam half-power width along the core--jet direction as the uncertainty of the component separations. The core flux density errors are assumed to be $10\%$, as they are dominated by the VLBI absolute flux density calibration uncertainty \citep{2021ApJ...919...40P}. In Fig.~\ref{1508:separation}, the separation of the jet from the core is displayed as a function of time. We fit a linear function to the radial proper motion of the jet component and obtain $0.117 \pm 0.078$\,mas\,yr$^{-1}$. The light curve constructed from the fitted flux densities of the core component is shown in Fig.~\ref{1508:lightcurve}. It seems that the source underwent a period of remarkable brightening, by more than a factor of 2, in 2017--2019, signaling the rising part of an outburst.

Coincidentally, the fitted displacement rate of the southern jet component ($0.117$\,mas\,yr$^{-1}$) is similar in magnitude to the declination component of the apparent astrometric proper motion ($0.147$\,mas\,yr$^{-1}$ southward) within the uncertainties. However, such a weak jet component with fitted flux density of $\sim 10-20$\,mJy, compared to the compact core with $\sim 100-300$\,mJy (Fig.~\ref{1508:lightcurve}), is unlikely to significantly influence the absolute astrometric position of the entire radio source because the jet component accounts for less than $\sim 10\%$ of the total emission. Therefore the displacement rate of the centroid of mas-scale radio emission cannot be larger than about one-tenth of the jet component proper motion. The small changes in right ascension and declination of 1508$+$572 we observed in 2017--2020 (Fig.~\ref{fig:1508}) could rather be related to the on-going flux density outburst (Fig.~\ref{1508:lightcurve}) which might be caused by a newly ejected bright inner jet component that is, however, still blended with the core because of the limited angular resolution. Continuing VLBI imaging of 1508$+$572 could be able to resolve this new component within a couple of years. Here we restricted the study of the mas-scale radio properties of this source to the time interval covered by our astrometric monitoring only. A more detailed, astrophysically motivated analysis extending to longer periods is beyond the scope of this paper.

\subsection{2101$+$600}

The quasar 2101$+$600 ($z = 4.575$) was identified by \citet{2004ApJ...609..564S}, but was not studied in detail at optical wavelengths. In contrast, it was actively observed in radio with VLBI \citep{2002ApJS..141...13B,2008AJ....136..580P,2018A&A...618A..68F,2021MNRAS.507.3736Z}, and was found to be a gigahertz-peaked spectrum (GPS) source with its radio spectrum peaking near 1~GHz \citep{2017MNRAS.467.2039C}.

\begin{figure}[h]
\centering
 \includegraphics[width=0.4\textwidth, clip=, angle=0]{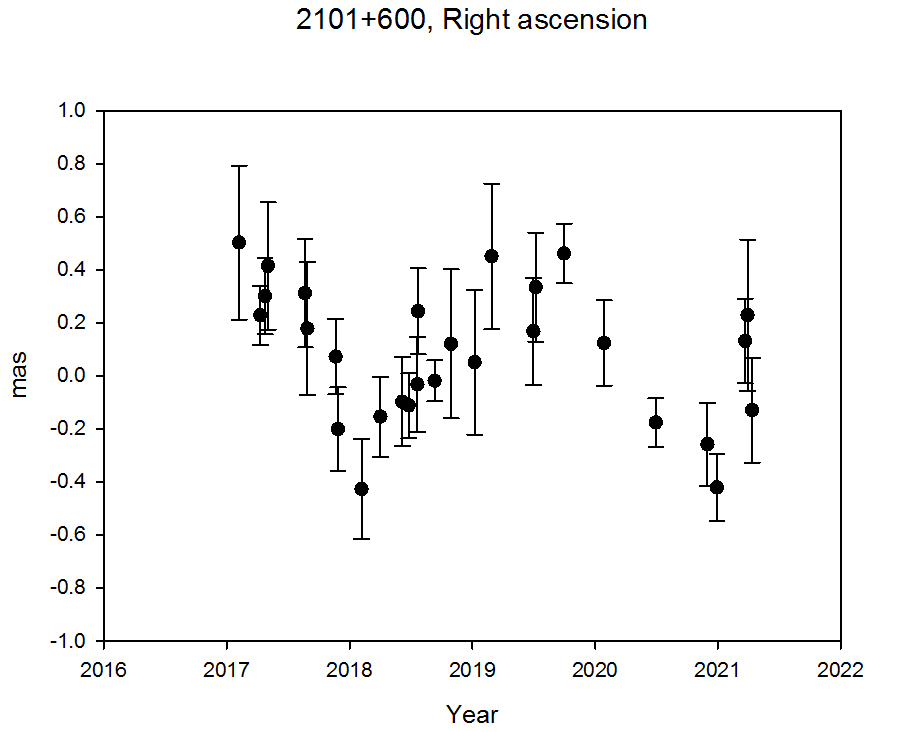}
 \includegraphics[width=0.4\textwidth, clip=, angle=0]{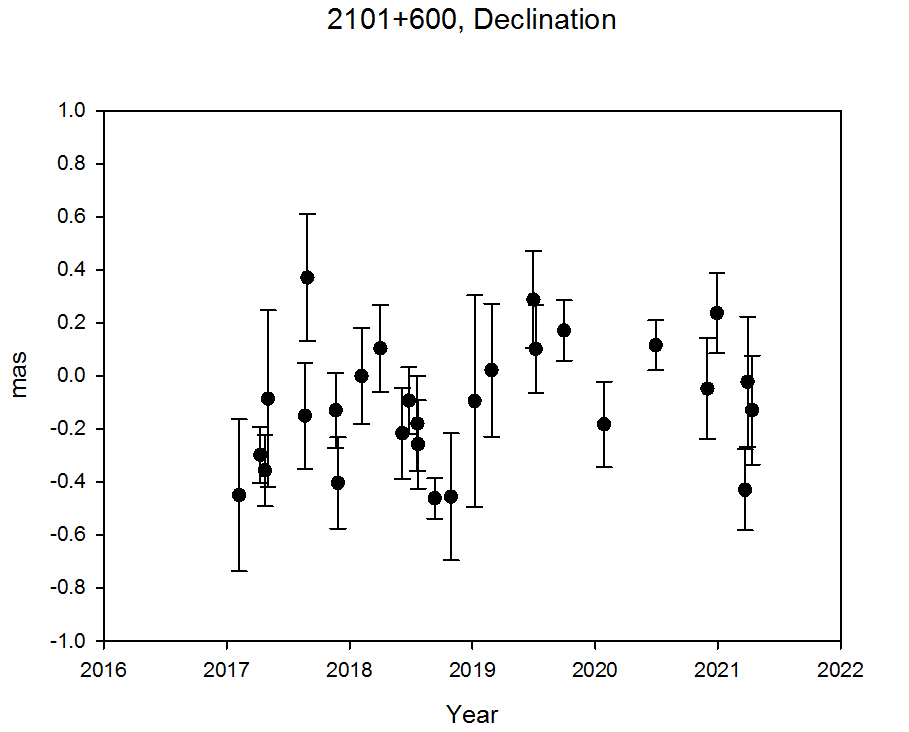}
 \caption{Relative astrometric coordinates (right ascension and declination) of 2101$+$600 as a function of time, between 2017 and 2020. The reference point corresponds to the solution aus2020a.crf. The error bars show $\pm 1\sigma$ formal uncertainty.} 
\label{fig:2101}
\end{figure}

\begin{figure}[h]
\centering
 \includegraphics[width=0.5\textwidth, trim={0 0 0 0}, clip=, angle=0]{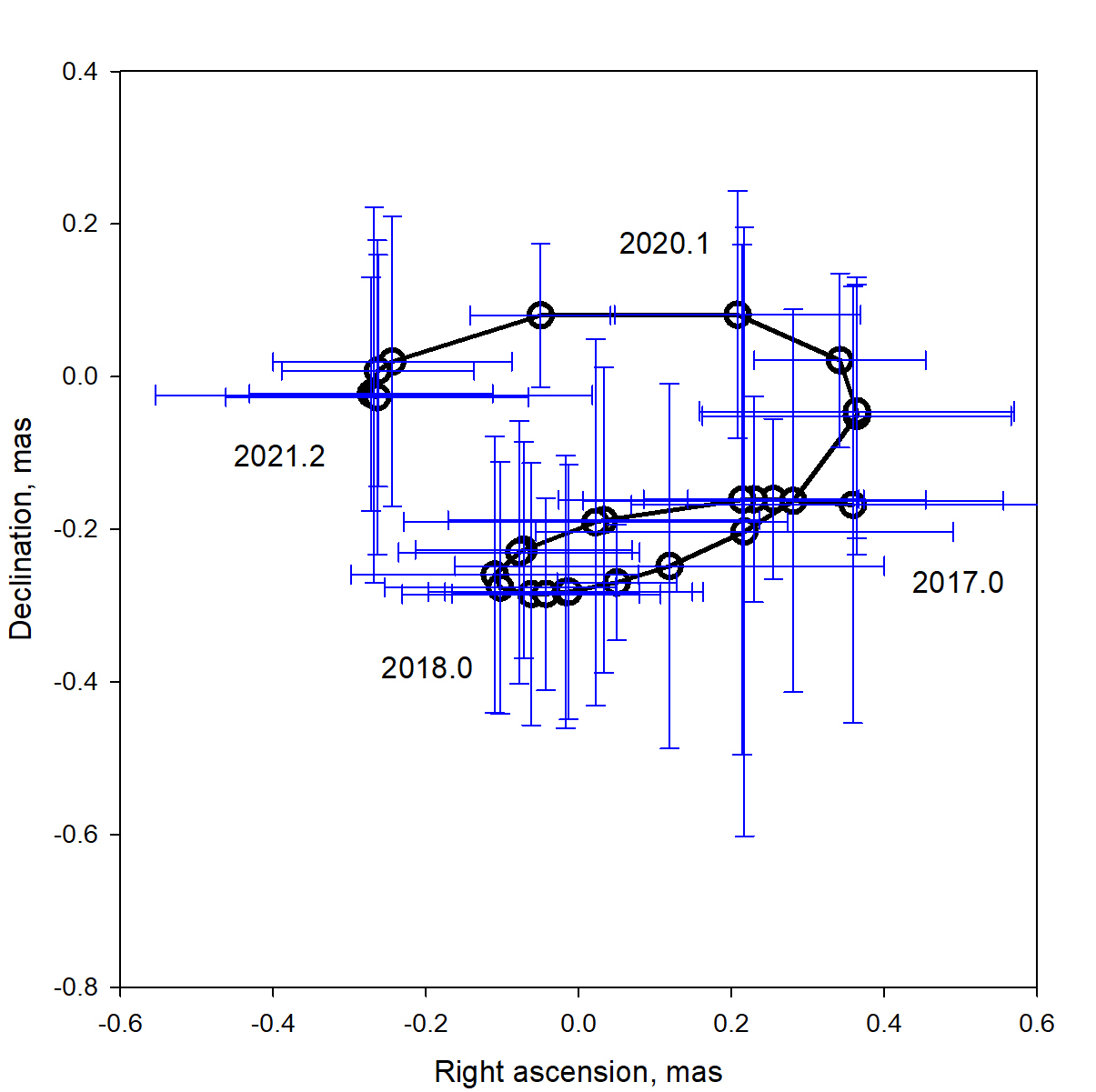}
\caption{Evolution of the position of the source 2101$+$600 in the sky plane over the period from 2017.0 to 2021.2.
The right ascension and declination time series (Fig.~\ref{fig:2101}) were fitted by the linear trend and harmonic signal with the period of 3 years. This approximation is shown by the black curve. The estimates of the amplitude of variations are $0.28 \pm 0.05$\,mas in right ascension and $0.11 \pm 0.05$\,mas in declination. The reference point corresponds to the solution aus2020a.crf. The error bars are from Fig.~\ref{fig:2101}, but referred to the fitting curve here.} \label{fig:2101-harmonic}
\end{figure}

Our VLBI coordinate time series in right ascension and declination are shown in Fig.~\ref{fig:2101}.
A quasi-periodic variation along the right ascension with period about 3 years was found for 2101$+$600. Fitting the time series in the right ascension yields the amplitude $0.28 \pm 0.05$\,mas. Taking this into account, a new estimate of the linear proper motion component along right ascension is $-0.55 \pm 0.25$\,mas\,yr$^{-1}$, i.e. about $20\%$ smaller than the estimate without accounting for the quasi-harmonic variation (Table~\ref{table_pm}). Longer-term observations to cover at least two suspected periods ($\ge 6$ yr) would be required to verify if this apparent periodicity persists. Although the time series in declination does not show the same systematic variations, we approximated both components with the 3-year harmonic. The corresponding approximation in sky plane is shown in Fig.~\ref{fig:2101-harmonic}.

VLBI imaging \citep[e.g.][]{2018A&A...618A..68F,2021MNRAS.507.3736Z} shows a structure extended to $\sim 10$\,mas in the east--west direction, i.e. along the right ascension axis. The overall source emission is dominated by the eastern feature that can, however, be adequately modeled with not a single but 3 circular Gaussian brightness distribution components. 

The western feature is comparably weak, and the faintest component in between the eastern and western features could only be recently revealed with sensitive VLBA imaging at 8.4~GHz \citep{2021MNRAS.507.3736Z}. There is no change ($<0.04$\,mas\,yr$^{-1}$) detected in the separation of the eastern and western features over a time range of $\sim 13$\,yr. In contrast to a core--jet blazar with a relativistically beamed radio jet pointing towards the observer, \citet{2021MNRAS.507.3736Z} interpret the VLBI imaging data and the peaked overall radio spectrum as characteristic to a young compact symmetric object (CSO) with its bipolar jets misaligned with respect to the line of sight. In this scenario, the weak central component is the core, marking the physical centre of the AGN, similarly to the case of e.g. the bright low-redshift CSO PKS\,1117+146 \citep{1998MNRAS.297..559B}. Therefore, the quasar 2101+600 may be morphologically different from other high-redshift sources in our sample. If the CSO interpretation holds, the brightness peak is associated with a complex hotspot in the eastern lobe, rather than a compact synchrotron self-absorbed blazar core. It is the brighter eastern lobe which dominates the astrometric positioning of the source.

Figure~\ref{fig:2101-harmonic} shows the apparent celestial track of the astrometric solutions obtained in our project. The pattern of this track resembles that of a precessing jet. The phenomenon of mas-scale jet precession in quasars is well known \citep[see e.g.][and references therein]{2018MNRAS.478.3199B,2019A&A...630A.103B}. It is usually attributed to either the presence of a binary supermassive black hole (SMBH) system in the nucleus of the object or the Lense--Thirring effect in a viscous ``warped disc'' \citep[e.g.][]{2014MNRAS.441.1408T}.
 
For a binary nature of the nucleus in the source 2101$+$600, one might analyse a ``toy model'' consisting of two SMBHs, with masses $m_1$ and $m_2$. The third Kepler's law for such the binary system can be presented as 
 
\begin{equation}
 a = \left( \frac{G M T^2}{4\pi^2} \right)^{1/3},
\end{equation}

\noindent where $a$ is the semi-major axis of the elliptical orbit of the binary system of the total mass $M=m_1+m_2$, $T$ is its rest-frame period, and $G$ is the gravitational constant. Simple visual analysis of Fig.~\ref{fig:2101-harmonic} allows us to estimate the apparent period of precession as $T_{\rm app} \approx 3$~yr. Taking into account the cosmological time dilation, it translates into $T = T_{\rm app}/(1+z) \approx 0.54$~yr in the rest-frame of the quasar at $z=4.575$. Then the semi-major axis of the binary system can be estimated as 

\begin{equation}
    a = 0.32\times 10^{-2} \times \left( \frac{M}{10^9 \times M_{\odot}} \right)^{1/3} \times \left( \frac{T}{0.54\,\mathrm{yr}} \right)^{2/3} \,\,\, \mathrm{pc}.
\end{equation}

\noindent For an equal distribution of masses in the binary system, $m_1 = m_2 = 0.5\times M = 0.5 \times 10^9 M_{\odot}$, the semi-major axis would only be $a\simeq 0.25\times 10^{-2}$~pc, or $\simeq 50$ times the gravitational radius $r_\mathrm{g} = 2 G m_1 c^{-2} \approx 0.48 \times 10^{-4}$~pc, where $c$ is the speed of light.

We intentionally normalize the estimate of the semi-major axis $a$ for the high value of the nucleus mass to underline that such the binary system would be a source of appreciable gravitational waves \citep[e.g.][]{2019NewAR..8601525D}. In reality, the mass is likely to be lower. But in any case, the source 2101$+$600 is an interesting target for studies of the growth of SMBH in early cosmological epochs.
Further astrophysical analysis of this source is outside of the scope of this paper. However, given the apparent variations of the astrometric position of 2101$+$600, illustrated by Fig.~\ref{fig:2101}, this object should not be used as a reference source involved in the formation of reference frames.

\begin{table}[h]
\caption{Proper motion estimates of the four quasars with 1-$\sigma$ formal errors for all experiments.}

\vspace{3mm}
\begin{tabular}{l c c c}

\hline
Quasar   &  PM in RA & PM in Dec & Number of experiments\\
& mas\,yr$^{-1}$ & mas\,yr$^{-1}$  & \\
\hline 
\hline
0901$+$697 & $-0.023 \pm 0.023$ & $-0.066 \pm 0.023$ & 28 \\
1428$+$422 & $-0.098 \pm 0.120$ & $-0.334 \pm 0.173$ &  7 \\
1508$+$572 & $-0.121 \pm 0.051$ & $-0.147 \pm 0.040$ & 18 \\ 
2101$+$600 & $-0.066 \pm 0.036$ & $+0.079 \pm 0.035$ & 26 \\ 
\hline
\vspace{3mm}
\label{table_pm}
\end{tabular}
\end{table}

\subsection{More Detailed Analysis of the Positional Time Series}

The resultant position of a radio source is related to the position of the phase center as detected on the interferometer baseline. While for a point-like radio source the interferometer response is the same for all baseline lengths and orientations, the response is more complicated for an extended radio source \citep{1990AJ.....99.1309C}. Since all four radio sources presented in this study display an extended structure, the estimates of each radio source position depend on the set of baselines actually tracking the source. For the custom-scheduled RUA experiments, we attempted to observe the three radio sources (0901$+$697, 1508$+$572, and 2101$+$600) by all available stations in the same mode whenever it was possible. However, the estimates presented in Table~\ref{table_pm} may be affected by the network geometry. The longer baselines provide better angular resolution of the extended source structure, and, consequently, are more sensitive to the small-scale structural variability. On the other hand, the longer baselines have limited observing time compared to shorter baselines due to sky visibility limitations for non-circumpolar sources. Therefore, quantification of the impact of the long baselines on the final position estimates is a challenging task.

Table~\ref{table_pm_3} shows the estimates of the apparent proper motion for the three sources observed in 25 RUA experiments in two versions: for all baselines and for baselines shorter than $6000$~km only. (The fourth source, 1428$+$422 was rarely observed in RUA experiments, therefore it is not included in this analysis.) The 6000-km threshold cuts off the longest baselines (Ys--Bd, Ys--Sh, Zc--Sh, Sv--Sh). 
This effectively reduces the maximum resolution capability of the total RUA network, and suppresses the effect of the structure as close components remain unresolved at the shorter baselines. This makes the daily estimates of the radio sources positions internally more consistent.

The proper motions for 0901$+$697 and 2101$+$600 in the first column of Table~\ref{table_pm_3} are very close to the proper motions from Table~\ref{table_pm} because almost all experiments were done with the RUA network. The difference in the estimates is within the reported 3-$\sigma$ uncertainty. Exclusion of the long baselines from the subset of data leads to a marginal change in proper motions of 0901$+$697. The difference between 2101$+$600 proper motions in Columns 3 and 5 (Table~\ref{table_pm_3}) is more notable ($0.114$~mas) but it is still within the combined uncertainty ($0.132$~mas). One could conclude that baselines of different lengths are likely to produce different positions of the phase center for 2101$+$600, though the total effect on the proper motion based on the 4-yr period of time is negligible. We believe that the apparently precessing motion of the jet (Figure~\ref{fig:2101-harmonic}) plays a more important role for the analysis of the linear trend in the position of the radio source 2101$+$600.

\begin{table}[h]
\caption{Proper motion estimates of three quasars with 1-$\sigma$ formal errors.
Columns 2 and 3 -- from RUA experiments only, including all baselines. 
Columns 4 and 5 -- from RUA experiments only, for baselines shorter than $6000$~km}
\vspace{3mm}
\begin{tabular}{l c c c c c c c}

\hline
Quasar   &  PM in RA & PM in Dec & PM in RA &  PM in Dec & Number of \\
& mas\,yr$^{-1}$ & mas\,yr$^{-1}$  & mas\,yr$^{-1}$ & mas\,yr$^{-1}$ & experiments\\
\hline 
\hline
0901$+$697 & $-0.003 \pm 0.029$ & $-0.069 \pm 0.027$ &  $-0.028 \pm 0.023$ &  $-0.063 \pm 0.032$ & 25\\
1508$+$572 & $-0.243 \pm 0.086$ & $-0.341 \pm 0.131$ &  $-0.039 \pm 0.039$ &  $-0.132 \pm 0.093$ & 10\\ 
2101$+$600 & $-0.080 \pm 0.041$ & $+0.116 \pm 0.037$ &  $-0.027 \pm 0.031$ & $+0.002 \pm 0.024$ & 24\\ 
\hline
\vspace{3mm}
\label{table_pm_3}
\end{tabular}
\end{table}

\section{Summary and Conclusion}

\citet{1966ApJ...143..379K} showed that as the apparent proper motion induced by anisotropic expansion of the Universe may be proportional to the redshift or may be not, the primordial gravitational waves, if they exist in the early Universe, would cause a redshift-dependent systematic effect in the apparent proper motion. The $5/6$ of the systematic is to appear in second-order transverse vector spherical harmonics (quadrupole systematic pattern around the sky) equally distributed between the ``electric'' and ``magnetic'' components, and the minor fraction of the mean squared proper motion resides in the higher harmonics \citep{1996ApJ...465..566P, 1997ApJ...485...87G}. The second-order spherical harmonics were not found during the search among all radio sources in the past \citep[e.g.][]{1997ApJ...485...87G, 2011A&A...529A..91T, 2012A&A...547A..59M}, therefore, in this paper we focus on the most distant targets to detect potential anomalies. Any unusually large values of apparent proper motion of one of the high-redshift radio sources may hint that the primordial gravitational waves are strong enough to be detected.

The apparent proper motion in the positions of the reference radio sources due to the variations of the intrinsic brightness distribution is commonly within $0.1$\,mas\,yr$^{-1}$ \citep{2003A&A...403..105F} and rarely exceeds $0.5$\,mas\,yr$^{-1}$ \citep{2011A&A...529A..91T}. In an extreme scenario, the positional evolution with amplitude up to $1$\,mas\,yr$^{-1}$ was detected on a very short time scale, less than $1$~yr \citep{2007AstL...33..481T} due to, presumably, a brightening of a moving relativistic jet. However, this large value of motion in one direction is followed by a similar proper motion in opposite direction after fading out of the moving component, making the accumulated positional displacement is almost zero. The typical pattern of the positional variations over long period of time ($\ga 20$ years) is a piece-wise function made of interval length of 1--10 years \citep{2007AstL...33..481T}. As a result, the total apparent proper motion induced by the change in the intrinsic structure is suppressed over the whole period of observations. It also should be noted that the expansion of the Universe imposes a dilation factor $(1+z)^{-1}$, which results in the apparent ``slowing down'' of the processes in distant sources. Therefore, one could generally consider $0.1$\,mas\,yr$^{-1}$ as an upper limit on the linear drift for a typical radio source at $z\approx1$. Then, assuming that the systematic signal due to the primordial gravitational waves grows is proportional to the redshift, an apparent proper motion of $\ga 0.4$\,mas\,yr$^{-1}$, being persistent over several years, may be considered as anomalous at $z\approx4$, if no substantial structure of the given radio source is detected.
The goal of this study was to monitor the astrometric positions of high-redshift radio sources and to detect any anomalous linear drift in the apparent positions. Assuming the persistent linear drift has an amplitude of $0.5$\,mas\,yr$^{-1}$, the total displacement of a radio source during a 4-yr interval would reach 2~mas. 

The median redshift of sources observed with VLBI and involved in astrometric and geodetic studies is about $z=1$. Only a few of the objects in this sample (less than $1\%$) have redshift $z>4$. \citep[]{2009A&A...506.1477T, 2018ApJS..239...20M}, 

We monitored the four radio sources presented here (0901$+$697, 1428$+$422, 1508$+$572, and 2101$+$600) for more than four years using large VLBI facilities around the Earth, and added VLBA observations of the source 1508$+$572 during 2017--2018. 

We did not find any sign of unusually large apparent proper motion among the radio sources under study.
However, their high-resolution VLBI images known from the literature reveal an extended structure for all the objects. The estimates of apparent proper motion (Table~\ref{table_pm}) do not exceed their statistical $3\sigma$ threshold of significance, except for the declination component of the radio source 1508$+$572. These minor, mostly insignificant positional variations tend to align with the intrinsic structure revealed by VLBI images. 

The four frequently observed radio sources display extended, changing radio structure. The amplitudes of the estimated apparent proper motions in this study vary typically in a range between $2\sigma-3\sigma$ (Table~\ref{table_pm}). The linear trends in right ascension or declination may exceed the statistical $3\sigma$ threshold of significance, were the radio sources observed for an additional $2-3$\,yr. An apparent quasi-periodic variation was detected in the right ascension of 2101$+$600. Its existence could also be confirmed with longer-term astrometric observations. Finally, the apparent proper motion as well as individual variations of the intrinsic structure do not reveal extraordinary features that may be attributed to a cosmological origin. 

We did not find the hypothetical anomalies of the cosmological nature in the positional change of the four radio sources studied here at $4.3 \leq z\leq 5.5$. This is consistent with the conclusion by \citet{2022arXiv220207536M} based on the optical study of astrometric proper motions of 60410 quasars at redshifts $0.5 \leq z \leq 7.03$. Our sample presented here is too small to allow us to make a definitive conclusion. However, we note that dozens of other known AGN targets at $z \geq 4.0$ have been imaged more recently with VLBI \citep[see, e.g.,][]{2022ApJS..260...49K}. Those sources as well as yet to be identified even more distant AGN at redshifts beyond the current de facto frontier of the ``known'' Universe at around $z = 7$ might become legitimate targets for future astrometric studies of anomalous proper motions. 

\acknowledgments
The National Radio Astronomy Observatory is a facility of the National Science Foundation operated under cooperative agreement by Associated Universities, Inc. 

The recent VLBA experiments were run by the geodetic group of the US Naval Observatory to monitor the radio reference frame sources with ten VLBA antennas. The authors acknowledge use of the VLBA under the US Naval Observatory's time allocation. This work supports USNO's ongoing research into the celestial reference frame and geodesy.

The RUA experiments were organised by the Institute of Applied Astronomy of the Russian Academy of Sciences and made use of an ad-hoc VLBI network which consists of three 32-m radio telescopes Badary, Svetloe, and Zelenchukskaya together with the 25-m radio telescope Sheshan of the Shanghai Astronomical Observatory of the Chinese Academy of Sciences, and the 40-m radio telescope of Yebes Observatory (Instituto Geogr\'{a}fico National, Spain). 
Badary, Svetloe, and Zelenchukskaya radio telescopes are operated by the Scientific Equipment Sharing Center of the Quasar VLBI Network.

The European VLBI Network is a joint facility of independent European, African, Asian and North American radio astronomy institutes. The EVN observations presented in this paper have been conducted under the project code ET036.

The Long Baseline Array is part of the Australia Telescope National Facility (grid.421683.a) which is funded by the Australian Government for operation as a National Facility managed by CSIRO. This work was supported by resources provided by the Pawsey Supercomputing Centre with funding from the Australian Government and the Government of Western Australia.

We acknowledge the use of archival calibrated VLBI data from the Astrogeo Center database maintained by Leonid Petrov. 

We are grateful to Sergei Kurdubov from the Institute of Applied Astronomy RAS for a fruitful discussion.

SF was supported by the Hungarian National Research, Development and Innovation Office (OTKA K134213). 

We acknowledge with gratitude very useful comments provided by the anonymous reviewer of the manuscript of this paper.

This paper is published with the permission of the CEO of the Geoscience Australia.

\software{sked \citep{2010ivs..conf...77G}, SFXC \citep{2015ExA....39..259K}, PIMA \citep{2011AJ....142...35P}, OCCAM \citep[v6.3,][]{2004ivsg.conf..267T}, AIPS \citep{2003ASSL..285..109G}, Difmap \citep{1997ASPC..125...77S}}

\bibliography{main}{}
\bibliographystyle{aasjournal}



\end{document}